# A generalized model for Yang-Fourier transforms in fractal space

## Xiao-Jun Yang

Department of Mathematics & Mechanics, China University of Mining & Technology, Xuzhou, 221008, P. R. China

Email: dyangxiaojun@g163.com

**Abstract** – Local fractional calculus deals with everywhere continuous but nowhere differentiable functions in fractal space. The Yang-Fourier transform based on the local fractional calculus is a generalization of Fourier transform in fractal space. In this paper, local fractional continuous non-differentiable functions in fractal space are studied, and the generalized model for the Yang-Fourier transforms derived from the local fractional calculus are introduced. A generalized model for the Yang-Fourier transforms in fractal space and some results are proposed in detail.

## 1. Introduction

Local fractional calculus has been revealed a useful tool in areas ranging from fundamental science to engineering in the past ten years [1-10]. It is important to deal with the continuous functions (fractal functions), which are irregular in the real world. Recently, some model for engineering derived from local fractional derivative was proposed [10]. The Yang-Fourier transform based on the local fractional calculus was introduced [6] and Yang continued to study this subject [10]. The importance of Yang-Fourier transform for fractal functions derives from the fact that this is the only mathematic model which focuses on local fractional continuous functions derived from local fractional calculus. The Yang-Fourier transform may be of great importance for physical and technical applications, and its mathematical beauty makes it an interesting study for pure mathematicians as well [10-13]. Here, our attempt to model generalized Yang-Fourier transforms.

## 2. Preliminaries

### 2.1. Notations and recent results

**Definition 1**
If there exists the relation [10, 12-14]
$$|f(x) - f(x_0)| < \varepsilon^\alpha \tag{2.1}$$
with $|x - x_0| < \delta$, for $\varepsilon, \delta > 0$ and $\varepsilon, \delta \in \mathbb{R}$.
Now $f(x)$ is called local fractional continuous at $x = x_0$, denote by $\lim_{x \to x_0} f(x) = f(x_0)$. Then $f(x)$ is called local fractional continuous on the interval $(a, b)$, denoted by [10, 12, 13]

$$f(x) \in C_\alpha(a, b). \tag{2.2}$$

**Definition 2**
A function $f(x)$ is called a non-differentiable function of exponent $\alpha$, $0 < \alpha \leq 1$, which satisfy Hölder function of exponent $\alpha$, then for $x, y \in X$ such that [10, 12, 13]

$$|f(x) - f(y)| \leq C|x - y|^\alpha. \tag{2.3}$$

**Definition 3**
A function $f(x)$ is called to be continuous of order $\alpha$, $0 < \alpha \leq 1$, or shortly $\alpha$ continuous, when we have the following relation [10, 12, 13]

$$f(x) - f(x_0) = o\left((x - x_0)^\alpha\right). \tag{2.4}$$

**Remark 1.** Compared with (2.4), (2.1) is standard definition of local fractional continuity. Here (2.3) is unified local fractional continuity.

**Definition 4**
Setting $f(x) \in C_\alpha(a, b)$, local fractional derivative of $f(x)$ of order $\alpha$ at $x = x_0$ is defined by [4, 5, 7-9, 10, 12-14]

$$\begin{aligned} f^{(\alpha)}(x_0) &= \left.\frac{d^\alpha f(x)}{dx^\alpha}\right|_{x=x_0} \\ &= \lim_{x \to x_0} \frac{\Delta^\alpha(f(x) - f(x_0))}{(x - x_0)^\alpha} \end{aligned} \tag{2.5}$$

where $\Delta^\alpha (f(x)-f(x_0)) \cong \Gamma(1+\alpha)\Delta(f(x)-f(x_0))$.

For any $x \in (a,b)$, there exists [10, 12, 13]

$$f^{(\alpha)}(x) = D_x^{(\alpha)} f(x), \quad (2.6)$$

denoted by

$$f(x) \in D_x^{(\alpha)}(a,b). \quad (2.7)$$

**Definition 5**

Setting $f(x) \in C_\alpha(a,b)$, local fractional integral of $f(x)$ of order $\alpha$ in the interval $[a,b]$ is defined [4, 6, 10, 12-14]

$$\begin{aligned} {_aI_b^{(\alpha)}} f(x) &= \frac{1}{\Gamma(1+\alpha)} \int_a^b f(t)(dt)^\alpha \\ &= \frac{1}{\Gamma(1+\alpha)} \lim_{\Delta t \to 0} \sum_{j=0}^{j=N-1} f(t_j)(\Delta t_j)^\alpha \end{aligned} \quad (2.8)$$

where $\Delta t_j = t_{j+1} - t_j$, $\Delta t = \max\{\Delta t_1, \Delta t_2, \Delta t_j, ...\}$ and $[t_j, t_{j+1}]$, $j = 0,...,N-1$, $t_0 = a, t_N = b$, is a partition of the interval $[a,b]$.

Here, it follows that

$${_aI_a^{(\alpha)}} f(x) = 0 \text{ if } a = b \quad (2.9)$$

and

$${_aI_b^{(\alpha)}} f(x) = -{_bI_a^{(\alpha)}} f(x) \text{ if } a < b \quad (2.10)$$

For any $x \in (a,b)$, there exists

$${_aI_x^{(\alpha)}} f(x), \quad (2.11)$$

denoted by

$$f(x) \in I_x^{(\alpha)}(a,b). \quad (2.12)$$

**Remark 2**. If $f(x) \in D_x^{(\alpha)}(a,b)$, or ${_{x_0}I_x^{(\alpha)}}(a,b)$, we have

$$f(x) \in C_\alpha(a,b). \quad (2.13)$$

**2.2. Recent results**

Suppose that $f(x), g(x) \in D_\alpha(a,b)$, the following differentiation rules are valid [5,14]:

$$\frac{d^\alpha(f(x) \pm g(x))}{dx^\alpha} = \frac{d^\alpha f(x)}{dx^\alpha} \pm \frac{d^\alpha g(x)}{dx^\alpha}; \quad (2.14)$$

$$\frac{d^\alpha(f(x)g(x))}{dx^\alpha} = g(x)\frac{d^\alpha f(x)}{dx^\alpha} + f(x)\frac{d^\alpha g(x)}{dx^\alpha}; \quad (2.15)$$

$$\frac{d^\alpha\left(\frac{f(x)}{g(x)}\right)}{dx^\alpha} = \frac{g(x)\frac{d^\alpha f(x)}{dx^\alpha} + f(x)\frac{d^\alpha g(x)}{dx^\alpha}}{g(x)^2} \quad (2.16)$$

if $g(x) \neq 0$;

$$\frac{d^\alpha(Cf(x))}{dx^\alpha} = C\frac{d^\alpha f(x)}{dx^\alpha}; \quad (2.17)$$

if $C$ is a constant.

If $y(x) = (f \circ u)(x)$ where $u(x) = g(x)$, then

$$\frac{d^\alpha y(x)}{dx^\alpha} = f^{(\alpha)}(g(x))\left(g^{(1)}(x)\right)^\alpha. \quad (2.18)$$

**Theorem 1** [7,14]

Suppose that $f(x), g(x) \in C_\alpha[a,b]$, then

$${_aI_b^{(\alpha)}}[f(x) \pm g(x)] = {_aI_b^{(\alpha)}} f(x) \pm {_aI_b^{(\alpha)}} g(x). \quad (2.19)$$

**Theorem 2** [7,14]

If $f(x) = g^{(\alpha)}(x) \in C_\alpha[a,b]$, then we have

$${_aI_b^{(\alpha)}} f(x) = g(b) - g(a). \quad (2.20)$$

**Theorem 3** [7,14]

If $g(x) \in C_1[a,b]$ and $(f \circ g)(s) \in C_\alpha[g(a), g(a)]$. Then we have

$${_{g(a)}I_{g(b)}^{(\alpha)}} f(x) = {_aI_b^{(\alpha)}}(f \circ g)(s)[g'(s)]^\alpha. \quad (2.21)$$

**Theorem 4** [7, 14]

Suppose that $f(x), g(x) \in D_\alpha(a,b)$ and $f^{(\alpha)}(x), g^{(\alpha)}(x) \in C_\alpha[a,b]$. Then we have

$${_aI_b^{(\alpha)}} f(t) g^{(\alpha)}(t) = [f(t)g(t)]_a^b - {_aI_b^{(\alpha)}} f^{(\alpha)}(t) g(t). \quad (2.22)$$

**2.3. The Yang-Fourier transforms in fractal space**

**Definition 6**

Suppose that $f(x) \in C_\alpha(-\infty, \infty)$, the Yang-Fourier transform, dented by $F_\alpha\{f(x)\} \equiv f_\omega^{F,\alpha}(\omega)$, is written in the form [10, 12, 13]

$$\begin{aligned} F_\alpha\{f(x)\} &= f_\omega^{F,\alpha}(\omega) \\ &= \frac{1}{\Gamma(1+\alpha)} \int_{-\infty}^{\infty} E_\alpha(-i^\alpha \omega^\alpha x^\alpha) f(x)(dx)^\alpha \end{aligned} \quad (2.23)$$

where the latter converges.

And of course, a sufficient condition for convergence is

$$\left|\frac{1}{\Gamma(1+\alpha)} \int_{-\infty}^{\infty} f(x) E_\alpha(-i^\alpha \omega^\alpha x^\alpha)(dx)^\alpha\right| \\ \leq \frac{1}{\Gamma(1+\alpha)} \int_{-\infty}^{\infty} |f(x)|(dx)^\alpha < K < \infty. \quad (2.24)$$

**Definition 7**

If $F_\alpha\{f(x)\} \equiv f_\omega^{F,\alpha}(\omega)$, its inversion formula is written in the form [10, 12, 13]

$$f(x) = F_\alpha^{-1}\left(f_\omega^{F,\alpha}(\omega)\right): \qquad (2.25)$$

$$= \frac{1}{(2\pi)^\alpha} \int_{-\infty}^{\infty} E_\alpha\left(i^\alpha \omega^\alpha x^\alpha\right) f_\omega^{F,\alpha}(\omega)(d\omega)^\alpha, x > 0.$$

## 3. Motivation of the generalized Yang-Fourier transforms in fractal space

If $f(x)$ is $2l$-periodic and local fractional continuous on $[-l, l]$, we have

$$f(x) = \sum_{k=-\infty}^{\infty} C_n E_\alpha\left(\frac{\pi^\alpha i^\alpha (nx)^\alpha}{l^\alpha}\right), \qquad (3.1)$$

where its coefficients is

$$C_n = \frac{1}{(2l)^\alpha} \int_{-l}^{l} f(x) E_\alpha\left(\frac{-\pi^\alpha i^\alpha (nx)^\alpha}{l^\alpha}\right)(dx)^\alpha. \qquad (3.2)$$

Let us set $C_n = \frac{\Gamma(1+\alpha)}{(2l)^\alpha} C_n^t$. We have

$$f(x) = \frac{1}{(2l)^\alpha} \sum_{k=-\infty}^{\infty} C_n^t E_\alpha\left(\frac{\pi^\alpha i^\alpha (nx)^\alpha}{l^\alpha}\right), \qquad (3.3)$$

where its coefficients is

$$C_n^t = \frac{1}{\Gamma(1+\alpha)} \int_{-l}^{l} f(x) E_\alpha\left(\frac{-\pi^\alpha i^\alpha (nx)^\alpha}{l^\alpha}\right)(dx)^\alpha. \qquad (3.4)$$

If we define

$$k_n^\alpha = \left(\frac{\pi n}{l}\right)^\alpha, \qquad (3.5)$$

then we have

$$(\Delta k_n)^\alpha = (k_{n+1} - k_n)^\alpha = \left(\frac{\pi}{l}\right)^\alpha. \qquad (3.6)$$

It is convenient to rewrite

$$f(x) = \frac{1}{(2\pi)^\alpha} \sum_{k=-\infty}^{\infty} C_k E_\alpha\left(i^\alpha x^\alpha k_n^\alpha\right)(\Delta k_n)^\alpha \qquad (3.7)$$

$$= \frac{1}{(2\pi)^\alpha} \int_{-\infty}^{\infty} C_k E_\alpha\left(i^\alpha x^\alpha k_n^\alpha\right)(dk_n)^\alpha$$

as $l \to \infty$ and

$$C_k = \frac{1}{\Gamma(1+\alpha)} \int_{-\infty}^{\infty} f(x) E_\alpha\left(i^\alpha x^\alpha k_n^\alpha\right)(dx)^\alpha. \qquad (3.8)$$

*Case 1.*

Taking $k_n^\alpha = \omega^\alpha$ in (3.9) and (3.8), this leads to the following results

$$f(x) = \frac{1}{(2\pi)^\alpha} \int_{-\infty}^{\infty} C_k E_\alpha\left(i^\alpha x^\alpha \omega^\alpha\right)(d\omega)^\alpha \qquad (3.9)$$

and

$$C_k = \frac{1}{\Gamma(1+\alpha)} \int_{-\infty}^{\infty} f(x) E_\alpha\left(i^\alpha x^\alpha \omega^\alpha\right)(dx)^\alpha. \qquad (3.10)$$

**Remark 3.** The above are called the Yang-Fourier transform [10, 12, 13].

**Case 2.**

Taking $\omega^\alpha = (2\pi)^\alpha \omega'^\alpha$ in (3.9) and (3.8) implies that

$$f(x) = \int_{-\infty}^{\infty} C_k E_\alpha\left(i^\alpha x^\alpha \omega'^\alpha\right)(d\omega')^\alpha \qquad (3.11)$$

and

$$C_k = \frac{1}{\Gamma(1+\alpha)} \int_{-\infty}^{\infty} f(x) E_\alpha\left(i^\alpha x^\alpha \omega'^\alpha\right)(dx)^\alpha. \qquad (3.12)$$

**Case 3.**

Taking $\omega^\alpha = \frac{(2\pi)^\alpha}{\Gamma(1+\alpha)} \omega^{*\alpha}$, it follows from (3.9) and (3.8) that

$$f(x) = \frac{1}{\Gamma(1+\alpha)} \int_{-\infty}^{\infty} C_k E_\alpha\left(i^\alpha x^\alpha \frac{(2\pi)^\alpha}{\Gamma(1+\alpha)} \omega^{*\alpha}\right)(d\omega^*)^\alpha \qquad (3.13)$$

and

$$C_k = \frac{1}{\Gamma(1+\alpha)} \int_{-\infty}^{\infty} f(x) E_\alpha\left(-i^\alpha x^\alpha \frac{(2\pi)^\alpha}{\Gamma(1+\alpha)} \omega^{*\alpha}\right)(dx)^\alpha. \qquad (3.14)$$

**Definition 8 (Generalized Yang-Fourier transform)**

From (3.14) we get a generalized Yang-Fourier transform in the form

$$F_\alpha\{f(x)\} = f_\omega^{F,\alpha}(\omega) \qquad (3.15)$$

$$= \frac{1}{\Gamma(1+\alpha)} \int_{-\infty}^{\infty} f(x) E_\alpha\left(-i^\alpha h_0 x^\alpha \omega^\alpha\right)(dx)^\alpha$$

where $h_0 = \frac{(2\pi)^\alpha}{\Gamma(1+\alpha)}$ with $0 < \alpha \leq 1$.

A sufficient condition for convergence is

$$\frac{1}{\Gamma(1+\alpha)} \int_{-\infty}^{\infty} |f(x)|(dx)^\alpha < K < \infty. \qquad (3.16)$$

**Definition 9**
From (3.13) we get the inverse formula of the generalized Yang-Fourier transform in the form

$$F_\alpha^{-1}\left(f_\omega^{F,\alpha}(\omega)\right)$$
$$= f(x) \quad (3.17)$$
$$= \frac{1}{\Gamma(1+\alpha)} \int_{-\infty}^{\infty} f_\omega^{F,\alpha}(\omega) E_\alpha\left(i^\alpha h_0 x^\alpha \omega^\alpha\right) (d\omega)^\alpha$$

where $h_0 = \dfrac{(2\pi)^\alpha}{\Gamma(1+\alpha)}$ with $0 < \alpha \le 1$.

A sufficient condition for convergence is

$$\frac{1}{\Gamma(1+\alpha)} \int_{-\infty}^{\infty} \left|f_\omega^{F,\alpha}(\omega)\right| (d\omega)^\alpha < M < \infty. \quad (3.18)$$

## 4. Some results

The following formulas are valid:
$$F_\alpha\{af(x)+bg(x)\} = aF_\alpha\{f(x)\} + bF_\alpha\{g(x)\},$$
$$a,b \in C \quad (4.1)$$

$$F_\alpha\{f(x-c)\} = E_\alpha\left(i^\alpha c^\alpha x^\alpha\right) F_\alpha\{f(x)\},$$
$$c \in C \quad (4.2)$$

$$F_\alpha\{f(ax)\} = a^{-\alpha} f_\omega^{F,\alpha}(\omega/a), \, a>0 \quad (4.3)$$

$$F_\alpha^{-1}\{af_\omega^{F,\alpha}(\omega) + bg_\omega^{F,\alpha}(\omega)\}$$
$$= aF_\alpha^{-1}\{f_\omega^{F,\alpha}(\omega)\} + bF_\alpha^{-1}\{g_\omega^{F,\alpha}(\omega)\}, \quad a,b \in C$$
$$(4.4)$$

$$F_\alpha^{-1}\{f_\omega^{F,\alpha}(\omega+c)\} = f(x) E_\alpha\left(-i^\alpha c^\alpha x^\alpha\right), \, c \in C \quad (4.5)$$

$$F_\alpha\{f^{(\alpha)}(x)\} = -i^\alpha h_0 \omega^\alpha F_\alpha\{f(x)\}. \quad (4.6)$$

The above are proved in Appendix A.

**Theorem 5** *(Uniqueness of the generalized Yang-Fourier transforms)*
Let $F_\alpha\{f_1(x)\} = f_{\omega,1}^{F,\alpha}(\omega)$ and $F_\alpha\{f_2(x)\} = f_{\omega,2}^{F,\alpha}(\omega)$.
Suppose that $f_{\omega,1}^{F,\alpha}(\omega) = f_{\omega,2}^{F,\alpha}(\omega)$, then
$$f_1(x) = f_2(x). \quad (4.7)$$
*Proof.* Using the motivation of the generalized Yang-Fourier transforms yields the result.

**Definition 10**
The convolution of two functions, which satisfy the condition (3.16) and (3.18), is defined symbolically by

$$f_1(x) * f_2(x) = \frac{1}{\Gamma(1+\alpha)} \int_{-\infty}^{x} f_1(t) f_2(x-t) (d\ )^\alpha t.$$
$$(4.8)$$

As further results, the properties of the convolution of the non-differentiable functions for convenience read as:

The commutative rule:
$$f_1(x) * f_2(x) = f_2(x) * f_1(x); \quad (4.9)$$
The distributive rule:
$$f_1(x) * (f_2(x) + f_3(x)) = f_1(x) * (f_2(x) + f_3(x)). \quad (4.10)$$

**Theorem 6**
Suppose
that $F_\alpha\{f_1(x)\} = f_{\omega,1}^{F,\alpha}(\omega)$ and $F_\alpha\{f_2(x)\} = f_{\omega,2}^{F,\alpha}(\omega)$.
Then
$$F_\alpha\{f_1(x) * f_2(x)\} = f_{\omega,1}^{F,\alpha}(\omega) f_{\omega,2}^{F,\alpha}(\omega). \quad (4.11)$$
*Proof.* Taking into account the definitions of the convolution of two functions and the generalized Yang-Fourier transform implies that

$$F_\alpha\{f_1(x) * f_2(x)\}$$
$$= \frac{1}{\Gamma(1+\alpha)} \int_{-\infty}^{\infty} E_\alpha\left(i^\alpha h_0 x^\alpha \omega^\alpha\right) (f_1(x) * f_2(x)) (dx)^\alpha.$$

Successively, rearranging equation (4.11) becomes

$$\frac{1}{\Gamma(1+\alpha)} \int_{-\infty}^{\infty} \left(\frac{1}{\Gamma(1+\alpha)} \int_{-\infty}^{\infty} E_\alpha\left(-i^\alpha h_0 x^\alpha \omega^\alpha\right) f_2(x-t) (dx)^\alpha\right)$$
$$f_1(t) (dt)^\alpha$$
$$= \frac{1}{\Gamma(1+\alpha)} \int_{-\infty}^{\infty} E_\alpha\left(-i^\alpha h_0 t^\alpha \omega^\alpha\right) f_1(t) f_{\omega,2}^{F,\alpha}(\omega) (dt)^\alpha.$$
$$(4.12)$$

Take into account the relation
$$f_{\omega,2}^{F,\alpha}(\omega)$$
$$= \frac{1}{\Gamma(1+\alpha)} \int_{-\infty}^{\infty} E_\alpha\left(-i^\alpha h_0 (x-t)^\alpha \omega^\alpha\right) f_2(x-t) (d(x-t))^\alpha,$$
$$(4.13)$$
which follows from (3.12) that

$$f_{\omega,1}^{F,\alpha}(\omega) f_{\omega,2}^{F,\alpha}(\omega)$$
$$= \frac{1}{\Gamma(1+\alpha)} \int_{-\infty}^{\infty} E_\alpha\left(-i^\alpha h_0 t^\alpha \omega^\alpha\right) f_1(t) f_{\omega,2}^{F,\alpha}(\omega) (dt)^\alpha.$$
$$(4.14)$$

Hence we arrive at the result.
As a direct result, we have the following result.
**Theorem 7**
Let $F_\alpha\{f(x)\} = f_\omega^{F,\alpha}(\omega)$, then
$$\frac{1}{\Gamma(1+\alpha)} \int_{-\infty}^{\infty} |f(x)|^2 (dx)^\alpha = \frac{1}{\Gamma(1+\alpha)} \int_{-\infty}^{\infty} \left|f_\omega^{F,\alpha}(\omega)\right|^2 (d\omega)^\alpha.$$
$$(4.15)$$

*Proof.* Using the definition of convolution and inverse formula of generalized Yang-Fourier transform implies that

$$\overline{f(x)} = \frac{1}{\Gamma(1+\alpha)}\int_{-\infty}^{\infty} E_\alpha\left(i^\alpha h_0 \omega^\alpha x^\alpha\right) f_\omega^{F,\alpha}(\omega)(d\omega)^\alpha. \quad (4.16)$$

Furthermore

$$\frac{1}{\Gamma(1+\alpha)}\int_{-\infty}^{\infty} \overline{E_\alpha\left(i^\alpha h_0 \omega^\alpha x^\alpha\right)} \overline{f_\omega^{F,\alpha}(\omega)} (d\omega)^\alpha$$

$$= \frac{1}{\Gamma(1+\alpha)}\int_{-\infty}^{\infty} E_\alpha\left(-i^\alpha h_0 \omega^\alpha x^\alpha\right) \overline{f_\omega^{F,\alpha}(\omega)} (d\omega)^\alpha. \quad (4.17)$$

From (4.17), (4.16) becomes

$$\overline{f(x)} = \frac{1}{\Gamma(1+\alpha)}\int_{-\infty}^{\infty} E_\alpha\left(-i^\alpha h_0 \omega^\alpha x^\alpha\right) \overline{f_\omega^{F,\alpha}(\omega)} (d\omega)^\alpha. \quad (4.18)$$

Now we have

$$\frac{1}{\Gamma(1+\alpha)}\int_{-\infty}^{\infty} |f(x)|^2 (d x)^\alpha = \frac{1}{\Gamma(1+\alpha)}\int_{-\infty}^{\infty} f(x)\overline{f(x)}(d x)^\alpha. \quad (4.19)$$

Using (4.18) implies that

$$\frac{1}{\Gamma(1+\alpha)}\int_{-\infty}^{\infty} f(x)\overline{f(x)}(d x)^\alpha$$

$$= \frac{1}{\Gamma^2(1+\alpha)}\int_{-\infty}^{\infty} \overline{f_\omega^{F,\alpha}(\omega)} \left(\int_{-\infty}^{\infty} f(x) E_\alpha\left(-i^\alpha h_0 \omega^\alpha x^\alpha\right)(dx)^\alpha\right)(d\omega)^\alpha. \quad (4.20)$$

Successively, rearranging (4.20) yields

$$\frac{1}{\Gamma(1+\alpha)}\int_{-\infty}^{\infty} \overline{f_\omega^{F,\alpha}(\omega)} f_\omega^{F,\alpha}(\omega)(d\omega)^\alpha$$

$$= \frac{1}{\Gamma(1+\alpha)}\int_{-\infty}^{\infty} |f_\omega^{F,\alpha}(\omega)|^2 (d\omega)^\alpha. \quad (4.21)$$

Hence, the proof of theorem is completed.

## 5. Conclusions

In present paper we give a generalized Yang-Fourier transforms as follows:

$$F_\alpha\{f(x)\} = \frac{1}{\Gamma(1+\alpha)}\int_{-\infty}^{\infty} f(x) E_\alpha\left(-i^\alpha h_0 x^\alpha \omega^\alpha\right)(dx)^\alpha \quad (5.1)$$

and

$$f(x) = \frac{1}{\Gamma(1+\alpha)}\int_{-\infty}^{\infty} f_\omega^{F,\alpha}(\omega) E_\alpha\left(i^\alpha h_0 x^\alpha \omega^\alpha\right)(d\omega)^\alpha, \quad (5.2)$$

where $h_0 = \dfrac{(2\pi)^\alpha}{\Gamma(1+\alpha)}$ with $0 < \alpha \leq 1$.

The transforming functions are local fractional continuous. That is to say, it is fractal function defined on fractal sets. Fourier transforms in integer space are the special case of fractal dimension $\alpha = 1$. It is a tool to deal with differential equation with local fractional derivative.

## Appendix A.

Taking into account equation (2.19), we directly obtain formulas (4.1) and (4.4).
Now we start with equation (4.2).

$$F_\alpha\{f(x-c)\}$$

$$= \frac{1}{\Gamma(1+\alpha)}\int_{-\infty}^{\infty} f(x-c) E_\alpha\left(-i^\alpha h_0 x^\alpha \omega^\alpha\right)(dx)^\alpha$$

$$= \frac{1}{\Gamma(1+\alpha)}\int_{-\infty}^{\infty} E_\alpha\left(-i^\alpha h_0 c^\alpha \omega^\alpha\right) f(x-c)$$

$$E_\alpha\left(-i^\alpha h_0 (x-c)^\alpha \omega^\alpha\right)(d(x-c))^\alpha$$

(using (2.21))

$$= \frac{E_\alpha\left(-i^\alpha h_0 c^\alpha \omega^\alpha\right)}{\Gamma(1+\alpha)}\int_{-\infty}^{\infty} f(x) E_\alpha\left(-i^\alpha h_0 x^\alpha \omega^\alpha\right)(dx)^\alpha$$

$$= E_\alpha\left(i^\alpha h_0 c^\alpha \omega^\alpha\right) F_\alpha\{f(x)\}$$

Now we start with equation (4.3).

$$F_\alpha\{f(ax)\}$$

$$= \frac{1}{\Gamma(1+\alpha)}\int_{-\infty}^{\infty} f(ax) E_\alpha\left(-i^\alpha h_0 x^\alpha \omega^\alpha\right)(dx)^\alpha$$

$$= \frac{1}{a\Gamma(1+\alpha)}\int_{-\infty}^{\infty} f(ax) E_\alpha\left(-i^\alpha h_0 (ax)^\alpha \left(\frac{\omega}{a}\right)^\alpha\right)(d(ax))^\alpha$$

(using (2.21))

$$= \frac{1}{a\Gamma(1+\alpha)}\int_{-\infty}^{\infty} f(ax) E_\alpha\left(-i^\alpha h_0 (ax)^\alpha \left(\frac{\omega}{a}\right)^\alpha\right)(d(ax))^\alpha$$

$$= \frac{1}{a^\alpha} f_\omega^{F,\alpha}\left(\frac{\omega}{a}\right).$$

Now we start with equation (4.3).

$$F_\alpha\{f_\omega^{F,\alpha}(\omega+c)\}$$

$$= \frac{1}{\Gamma(1+\alpha)}\int_{-\infty}^{\infty} f_\omega^{F,\alpha}(\omega+c) E_\alpha\left(i^\alpha h_0 x^\alpha \omega^\alpha\right)(d\omega)^\alpha$$

$$= \frac{E_\alpha\left(-i^\alpha h_0 x^\alpha c^\alpha\right)}{\Gamma(1+\alpha)}\int_{-\infty}^{\infty} f_\omega^{F,\alpha}(\omega+c) E_\alpha\left(i^\alpha h_0 x^\alpha (\omega+c)^\alpha\right)$$

$$(d(\omega+c))^\alpha$$

(using (2.21))

$$= E_\alpha\left(-i^\alpha h_0 x^\alpha c^\alpha\right) f(x).$$

Now we start with equation (4.3).

$$F_\alpha\left\{f^{(\alpha)}(x)\right\}$$

$$= \frac{1}{\Gamma(1+\alpha)}\int_{-\infty}^{\infty} f^{(\alpha)}(x) E_\alpha\left(-i^\alpha h_0 x^\alpha \omega^\alpha\right)(dx)^\alpha$$

(using (2.20) and (2.2))

$$= f^{(\alpha)}(x) E_\alpha\left(-i^\alpha h_0 x^\alpha \omega^\alpha\right)\Big|_{-\infty}^{\infty}$$

$$- \frac{i^\alpha h_0 \omega^\alpha}{\Gamma(1+\alpha)}\int_{-\infty}^{\infty} f(x) E_\alpha\left(-i^\alpha h_0 x^\alpha \omega^\alpha\right)(dx)^\alpha$$

$$= -i^\alpha h_0 \omega^\alpha F_\alpha\{f(x)\}.$$

## References

[1] K. M. Kolwankar, A. D. Gangal, Fractional differentiability of nowhere differentiable functions and dimensions, Chaos, 6 (4) (1996) 505-513.

[2] A. Carpinteri, P. Cornetti, A fractional calculus approach to the description of stress and strain localization in fractal media, Chaos, Solitons and Fractals, 13 (2002) 85-94.

[3] F. B. Adda, J. Cresson, Quantum derivatives and the Schrödinger equation, Chaos, Solitons and Fractals, 19 (2004) 1323-1334.

[4] F. Gao, X. J.Yang, Z. X. Kang, Local fractional Newton's method derived from modified local fractional calculus, In Proceeding of the second Scientific and Engineering Computing Symposium on Computational Sciences and Optimization, pp.228-232, 2009.

[5] X. J. Yang, F. Gao, The fundamentals of local fractional derivative of the one-variable non-differentiable functions, World SCI-TECH R&D, 31 (5) (2009) 920-921(in Chinese).

[6] X. J. Yang, Z, X. Kang, C. H. Liu, Local fractional Fourier's transform based on the local fractional calculus, In Proceeding of The 2010 International Conference on Electrical and Control Engineering, pp.1242-1245, 2010.

[7] X. J. Yang, L. Li, R. Yang, Problems of local fractional definite integral of the one-variable non-differentiable function, World SCI-TECH R&D, 31 (4) (2009) 722-724 (in Chinese).

[8] X. J. Yang, Local fractional Laplace's transform based on local fractional calculus, Communications in Computer and Information Science, 153 (2011) 391-397.

[9] X. J. Yang, Applications of local fractional calculus to engineering in fractal time-space: Local fractional differential equations with local fractional derivative, ArXiv:1106.3010v1 [math-ph], 2011.

[10] W. P. Zhong, F. Gao, X. M. Shen, Applications of Yang-Fourier Transform to Local Fractional Equations with Local Fractional Derivative and Local Fractional Integral, Adv. Mat. Res., 461 (2012) 306-310.

[11] S. M. Guo, L. Q. Mei, Y. Li , Y. F. Sun, The improved fractional sub-equation method and its applications to the space–time fractional differential equations in fluid mechanics, Phys. Lett. A., 376 (4) (2011) 407-411.

[12] X. J. Yang, Generalized Sampling Theorem for Fractal Signals, Advances in Digital Multimedia, 1 (2) (2012) 88-92.

[13] X. J. Yang, Local fractional partial differential equations with fractal boundary problems, Advances in Computational Mathematics and its Applications, 1 (1) (2012) 60-63.

[14] X. J. Yang, A short note on local fractional calculus of function of one variable, Journal of Applied Library and Information Science, 1 (1) (2012) 1-13.